\documentclass[conference]{IEEEtran}

\usepackage{graphicx}
\usepackage{epstopdf}
\usepackage{graphics}

\usepackage{listings}

\usepackage{color}
\definecolor{gray}{rgb}{0.4,0.4,0.4}
\definecolor{darkblue}{rgb}{0.0,0.0,0.6}
\definecolor{cyan}{rgb}{0.0,0.6,0.6}

\usepackage{amssymb}
\usepackage{multirow}

\usepackage[hang,flushmargin]{footmisc}

\usepackage[finalnew]{trackchanges}
\soulregister\ref7
\soulregister\cite7
\addeditor{m}
\addeditor{f}

\ifCLASSINFOpdf
\else
\fi

\usepackage{array}

\begin{document}
\title{Blockchain-based Supply Chain Traceability:\\
	Token Recipes model Manufacturing Processes}

\author{\IEEEauthorblockN{Martin Westerkamp,
Friedhelm Victor and
Axel K\"upper}
\IEEEauthorblockA{Service-centric Networking\\ Telekom Innovation Laboratories, Technische Universit\"at Berlin\\
Berlin, Germany\\
\{westerkamp, friedhelm.victor, axel.kuepper\}@tu-berlin.de}
}

\let\oldtextbf\textbf
\renewcommand{\textbf}[1]{\mbox{\oldtextbf{#1}}}

\maketitle

\begin{abstract}
Growing consumer awareness as well as manufacturers' internal quality requirements lead to novel demands on supply chain traceability.
Existing centralized solutions suffer from isolated data storage and lacking trust when multiple parties are involved. Decentralized blockchain-based approaches attempt to overcome these shortcomings by creating digital representations of physical goods to facilitate tracking across multiple entities. However, they currently do not capture the transformation of goods in manufacturing processes. Therefore, the relation between ingredients and product is lost, limiting the ability to trace a product's provenance.
We propose a blockchain-based supply chain traceability system using smart contracts. 
In such contracts, manufacturers define the composition of products in the form of recipes. Each ingredient of the recipe is a non-fungible token that corresponds to a batch of physical goods. When the recipe is applied, its ingredients are consumed and a new token is produced. This mechanism preserves the traceability of product transformations.
The system is implemented for the Ethereum Virtual Machine and is applicable to any blockchain configuration that supports it.
Our evaluation reveals that the gas costs scale linearly with the number of products considered in the system. This leads to the conclusion that the solution can handle complex use cases.
\end{abstract}

\IEEEpeerreviewmaketitle

\section{Introduction}

Providing traceability of goods from resource to retailer has become increasingly important in the past decade.
Consumers have a larger interest in consuming goods that comply with certain ecological and ethical standards~\cite{Dabbene2014}.
Global supply chains have become complex to a greater extent, hampering quality management in manufacturers' procurement~\cite{Gualandris2015}.
Furthermore, regulations, international standardizations and increased consumer awareness imply novel requirements towards supply chain management systems.
For instance, the European Parliament postulates the traceability of food, requiring food suppliers and market actors to provide information about provenance of goods~\cite{eu178}.
In addition, the ISO 9001:2015 standard instructs organizations to monitor identifiability and traceability of products and services.\par%

To cope with these requirements, supply chain management systems should ideally be operated by multiple business partners to trace a product's origin and its transformation process.
However, traditional supply chain management systems are operated isolated from other participants and are not capable of providing comprehensible provenance information~\cite{Appelhanz2016}. Resulting shortcomings include insufficient trust between parties, isolated data storage and unsatisfactory standardization in communication and data formats~\cite{Abeyratne2016,korpela2017}.

Recently, blockchain technology has been proposed for providing enhanced traceability in supply chains~\cite{korpela2017,glaser2017,wuest2017,Kim2016,Toyoda2017}.
Major drivers originate from typical blockchain characteristics such as decentralization, verifiability and immutability that could tackle the observed shortcomings.\par%

Current blockchain-based solutions for supply chain traceability promote tracking goods over multiple tiers by utilizing markers such as RFID and QR codes~\cite{Abeyratne2016}.
This linkage mechanism enables proving provenance for anti-counterfeit with regard to high value goods such as diamonds\footnote{https://www.everledger.io/}, medicine~\cite{Hackius2017} or generally in the post-retail supply chain~\cite{Toyoda2017}.
But these approaches are limited to non-modifiable goods and do not consider the production processes.
Thus, it is neither possible to track a product after it has been processed, nor to trace an end-product's inputs towards its primary resources.\par
In contrast to existing solutions, we foster a representation mechanism for the convertibility of products. Instead of only projecting physical goods onto the blockchain in the form of tokens, our target is to document their transformation in the production process on the ledger. We therefore propose a set of smart contracts that handles modifiable goods by capturing their creation, transformation and exchange on a distributed ledger. As a result, not only the good's origin is traceable but also its inputs. Hereby, we tackle two requirements for supply chain management that conventional systems cannot deliver comprehensively~\cite{BECHINI2008342}. Firstly, customers are empowered to review a product's and its ingredients' quality in various dimensions such as environmental and labour standards. Secondly, in the context of quality management, the proposed system enables manufacturers to monitor the supply chain for multiple tiers rather than relying on information provided by suppliers. Our key contributions in this paper are as follows:
\begin{enumerate}
  \item We design a supply chain traceability system that models manufacturing processes as token recipes.
  \item We present a prototypical implementation for the Ethereum Virtual Machine using smart contracts.
  \item We evaluate our implementation with respect to smart contract execution costs and scalability with increasing product complexity in terms of inputs.
\end{enumerate}

\section{Background and related work}

\subsection{Conventional Supply Chain Provenance Systems}
The state of the art for tracing and tracking products throughout a supply chain is storing records of suppliers and customers in a centralized manner with each participant's supply chain management system~\cite{Appelhanz2016}.
This information is isolated or shared among suppliers and customers to achieve comprehensive insights on supply chain provenance~\cite{BECHINI2008342}.

In the context of supply chain management, traceability promotes following a good's path upwards to its origin, while tracking refers to the downward operation, from raw materials to end products~\cite{Dabbene2014}.
For storing, sharing and managing relevant information, a plethora of information systems is used in practice.
Warehouse and transport management systems focus on internal warehouse operations, while enterprise resource planning (ERP) systems include supplier management, reordering and billing~\cite{Gunasekaran2004}.
Dedicated supply chain management software applications target forecasting future demands and fulfilling these from available suppliers \cite{Helo2005}.
\change[m]{The introduction of novel identifier technologies such as RFID or QR codes enables  projecting physical goods onto digital systems}
{As no global view on the supply chain is given, provenance information is retrieved from the next tier, requiring trust in the supplier or a third party.
However, enabling multi-tier traceability is essential for decreasing risks caused by quality fluctuations in source products~\cite{Tse2011}.

To ensure the coupling between physical goods and digital representations, identification mechanisms are fundamental. 
The ISO 28219:2017 standard defines guidelines for creating globally valid identifiers that are enforced by utilizing bar codes or two-dimensional symbols such as QR codes or alternative RFID tags for projecting physical goods onto digital systems}~\cite{Costa2013}. 
\change[m]{To ensure a certain quality degree, \textsc{Bechini et al.} propose defining batches of goods which hold a multitude of quality features~\cite{BECHINI2008342}. Varying the batch size permits influencing the granularity to which products are traceable.}{Identifiers either refer to single goods or product batches. Varying the batch size permits influencing the granularity to which products are traceable. To ensure a certain quality degree, Bechini et al. propose defining batches of goods that are linked to a multitude of quality features~\cite{BECHINI2008342}.}

\add[m]{While traditional supply chain information systems are capable of uniquely identifying products, traceability is limited. This is mainly due to isolated data that reflects organizations' sourcing and sales. Tracing ingredients over multiple tiers would require shared data that is tamper proof while maintaining high accessibility.}

\subsection{Smart Contracts and the Ethereum Virtual Machine}
\label{subsec:smartcontracts}
\add[m]{Blockchains provide a distributed, shared state all participants agree on using a consensus algorithm. Their tamper proof characteristics facilitate the opportunity of introducing a global view on multi-tier supply chains \cite{Yuan2016}.}
To apply business logic in blockchains, smart contracts have been introduced. First proposed by Nick Szabo in 1994~\cite{szabo1994smart}, smart contracts are computer programs that enforce rules without requiring a third party. In the Bitcoin blockchain, a basic version of smart contracts is implemented through the means of a scripting system that facilitates use cases like multi-user accounts (multi-signature wallets) and escrow services.
The main technology advancement of the Ethereum blockchain~\cite{Buterin2014} is the introduction of a general purpose and turing complete smart contract system which is manifested in the Ethereum Virtual Machine (EVM).

In the EVM, program code is executed by miners and other network participants who verify state changes. A smart contract in Ethereum is typically written in a high level programming language, like Solidity or Viper, and compiled to bytecode that is then deployed on the blockchain.
Execution requires sending a transaction to the contract's address, while specifying which function is to be called given a set of parameters. These functions in turn can call other smart contracts if they have been programmed to do so.
Computationally intensive routines, however, are not suitable, as the execution has a cost attached to it. 
For each operation supported by the EVM, a gas cost is defined in the Ethereum yellow paper~\cite{Wood2014}, where the main cost drivers are operations that store or change values on the blockchain. A transaction therefore needs to contain a sufficient amount of gas in order to guarantee successful execution. The actual costs of a transaction depend on how much gas is needed, and a gas price a user is willing to spend for each unit of gas. The transaction costs spent by a user are awarded to the miner that includes the transaction in a new block, as s/he has to verify and execute the transaction.

\change[m]{Originally, the EVM was only part of the public Ethereum blockchain, but compatibility has been introduced for Bitcoin through Counterparty\footnote{https://counterparty.io/} by adding a secondary layer or Qtum\footnote{https://qtum.org/} which is a fork of Bitcoin, implementing the EVM on top of a UTXO rather than an account model.}{While the EVM was originally designed for the Ehereum blockchain, several projects aim to port it to other ledgers. For instance, Counterparty\footnote{https://counterparty.io/} adds a secondary computation layer to run EVM code on top of the bitcoin network. Qtum\footnote{https://qtum.org/} supplies a bitcoin fork implementing the EVM with the goal of abstracting from Ethereum's account based model in order to support bitcoin-like light clients.} Additionally, smart contracts for the EVM can also be run on permissioned blockchains like Quorum\footnote{https://github.com/jpmorganchase/quorum} or Hyperledger Burrow\footnote{https://www.hyperledger.org/projects/hyperledger-burrow}, removing the need to operate in a public environment.
To provide a common form of tokens in Ethereum, a standard interface referred to as \emph{Ethereum Request for Comments (ERC) 20} was introduced\footnote{https://github.com/ethereum/EIPs/blob/master/EIPS/eip-20.md}. 
Popular Ethereum wallets like Mist\footnote{https://github.com/ethereum/mist} and Parity\footnote{https://github.com/paritytech/parity/wiki/Tokens} support tokens that implement this interface out of the box.
All standard operations such as receiving balances and transferring tokens to other addresses are available in the wallet's GUI.

\add[m]{In contrast to ERC20 tokens which cannot be distinguished, a standard for non-fungible tokens, ERC721, is proposed for handling deeds\footnote{https://github.com/ethereum/EIPs/pull/841}. The target is to create a standardized interface for creating and trading discriminable tokens reflecting digital or physical goods. Consuming tokens or defining conditions for their generation is not included in the proposal, hampering its utilization for projecting production processes onto the blockchain. In order to allow for goods to be substituted by equal ones in a production process, a measure of comparability of tokens needs to be available. As ERC20 is strictly fungible and ERC721 is strictly non-fungible, the two standards do not offer a middle ground. }

\subsection{Blockchain-based Supply Chain \change[m]{Provenance}{Traceability}}
\label{subsec:RelatedBlockchainSupplyChain}
Several blockchain-based solutions have been proposed in literature to overcome current obstacles in guaranteeing a certain product quality \cite{vanDorp2003}, compliance with legal obligations \cite{BECHINI2008342} or to counter fraud \cite{Feng2017}.\par
While there are various approaches tackling supply chain traceability, they are mostly concerned with tracing single, non-modifiable goods\footnote{https://www.ascribe.io/}~\cite{feng2016,Kim2016}. In fact, they target proving a good's authenticity and ownership via multiple hops. For instance, Kim and Laskowski present an ontology that promotes provenance in supply chains using the Ethereum blockchain \cite{Kim2016}. The proposed contract is implemented using Solidity and supports several typical supply chain operations such as \textit{produce} and  \textit{consume}. Nevertheless, the definition of novel functions and properties in the contract is limited to a rigid type system and the production of new goods out of existing resources is not possible.\par
A common issue in supply chain traceability is the projection of physical goods onto a digital representation.
Tian suggests a token representation that implies benefits such as proving authenticity by attaching RFID chips or QR codes to link physical products with their digital counterparts on the blockchain~\cite{feng2016}.
It inherently allows for proving ownership and permits transferring it to another party.
Hereby, a good can be tracked from creation to retail.
Toyoda et al. propose a blockchain-based post-retail supply chain system for anti counterfeits in which they utilize this linkage to enable tracing products after being sold in retail~\cite{Toyoda2017}.\par
Reviewing the here presented approaches for providing traceability in the supply chain unveiled a major shortcoming for more complex use cases. While they permit creating digital representations of physical goods which facilitate tracking across multiple entities, this connection is lost in case the product is processed.


\begin{figure*}[ht]
	\centering
	\includegraphics[width=1\textwidth]{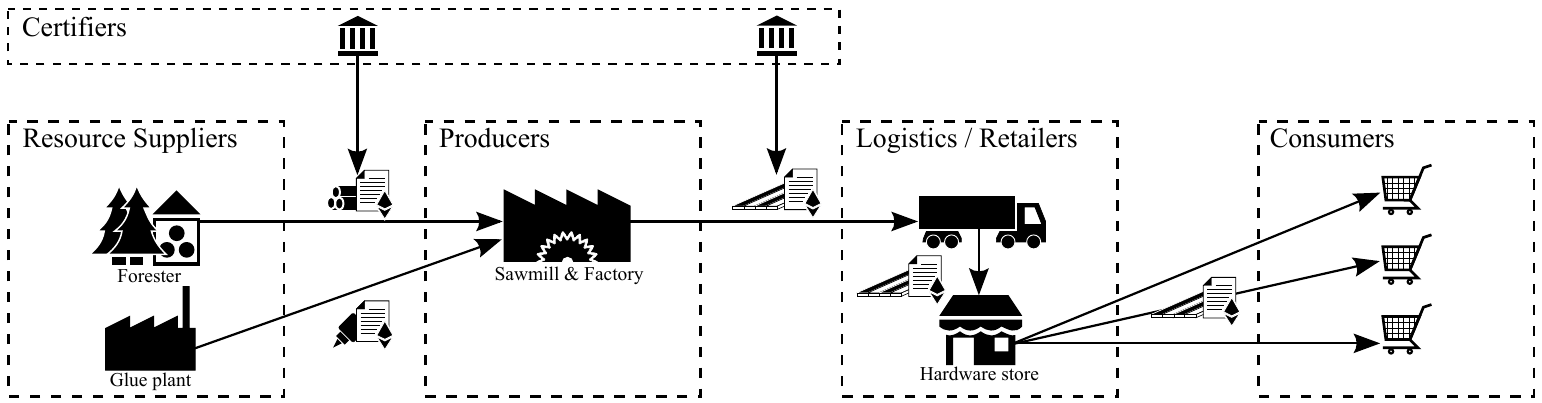}
	\caption{Projecting the production process onto digital tokens using smart contracts in an example supply chain}
	\label{fig:concept_supply_chain}
\end{figure*}

\section{Concept}
\label{concept}

We propose a blockchain-based, decentralized supply chain management system based on smart contracts. In order to provide comprehensive provenance information to consumers and producers, we maintain the relationship between resources and products in manufacturing processes. The concept is based on two key ideas: representing physical goods in the form of digital tokens, and recipes that enable their transformation. Additional functionality like certifying goods, transferring, splitting and combining tokens facilitates cross-business traceability. The following subsections describe each of these aspects in detail and introduce the participant's roles based on a specific example\add[m]{ visualized in Figure \ref{fig:concept_supply_chain}}.

\subsection{Tokenization of goods}

For each type of good managed in the supply chain, a smart contract is set up. Within such a smart contract, tokens can be created that represent physical goods.
One token corresponds to one batch of goods that could be measured in items, weight, volume or size.
The batch size is flexible, so that large quantities of goods, but also single entities are manageable.
The tokens are non-fungible, meaning that each token is unique. This allows distinguishing between batches of the same type of good.
To apply this concept, after manufacturing or sourcing a batch of products in the physical world, the contract owner creates digital tokens.

\subsection{Recipes for good transformation}
In order to digitally project a manufacturing process, several tokens can be transformed into a new token.
When creating a new smart contract, the product composition is defined. Comparable to a recipe, the creator specifies a number of input goods and corresponding amounts that are required for the creation a new product.
Following the recipe, when a batch of goods is to be created, the owner of the contract needs to possess the required input goods in sufficient amounts. In fact, the specific batches of input tokens need to be specified, so that they can be consumed by the smart contract automatically. If a batch of units is not entirely depleted, the remaining units are kept so that they can be used for future manufacturing.
Only contract owners can generate token batches, as a contract always corresponds to a specific producer's or supplier's product only s/he produces.
However, token owners can split, merge, transfer and consume batches.

\subsection{Certified goods}
Some goods are equally usable in a manufacturing process, because they are equal in type. They could also be equal in terms of compliance with standards indicating similar quality. However, the goods may originate from different sources. With the help of a certificate contract, multiple token contracts can be defined to be equal. This introduces an ontology for defining product inputs during token creation. In this aspect, our approach bears similarity with the architecture proposed by Bechini et al. \cite{BECHINI2008342}.
The certificates can be used in place of a specific good when defining a recipe as part of a manufacturing process. Once a batch is to be created, the input tokens are checked on whether they conform to the specified certificate.

\subsection{Roles}
A set of roles emerge from the actions participants can perform in the system.
A user may act in a single or multiple roles. Figure~\ref{fig:concept_supply_chain} illustrates these roles based on a simplified example taken from the wood processing industry. It contains participating entities as icons, surrounded by dashed lines separating the roles. Edges represent interaction, such as a token transfer or a certification. Smart contract symbols attached to every good indicate that a smart contract exists on a blockchain representing batches of it as digital tokens.
We identify the following roles that are also summarized in \mbox{Table \ref{table:roles}}.\newpage

\begin{enumerate}
	\item \textit{Resource suppliers}
	create goods without any input. When a forester cuts trees, several inputs, such as the labour or gasoline for machines, are required. However, they never become physical part of the resulting product. As every supply chain will only have a partial view of the real world, resource suppliers that do not consider all inputs must exist in the model. In our example, the glue plant is also a resource supplier as it has no inputs, but in other scenarios its inputs may be modeled to originate from another entity. A resource supplier may split or merge batches before they are transferred to other entitites. In the example, the forester transfers batches of logs to the sawmill.

	\item \textit{Producers} in turn do require input goods for producing outputs. To represent this process on the blockchain, the processing industry entity first defines all the goods that are required to create a certain product. The input goods are acquired physically, while this process is digitally captured through token transfers. Consequently, the producer owns the input tokens which are required for the creation of a new token. In the example, the sawmill is a producer that acquires batches of logs as well as glue. To ensure that sufficient input tokens are present, it can merge batches. The sawmill is then able to produce a batch of edge-glued wood, which it could split and transfer to a logistics or retail company.

	\item \textit{Logistics} and \textit{Retail} firms acquire tokens, but do not alter a good itself.
	For example, a batch of 100 units of edge-glued wood could be split, transferred to a logistics company and distributed to multiple hardware stores. If goods have been obtained by a wholesaler, merged batches could be transferred to other retailers. Individual customers could also keep ownership without consuming the item, in order to resell it to someone else. In this scenario, the customer is not strictly modelled as a consumer.

	\item \textit{Consumers} ultimately receive and consume products. This process results in the representing token batch to be deleted, meaning it can no longer be used as part of the supply chain. However, its provenance, can still be verified.
	If the retail customer is not modelled as part of the supply chain, a hardware store may act as a consumer: it removes the token from the supply chain when the item has been sold.

	\item \textit{Certifiers} issue certificates of equality for multiple goods. They are responsible for guaranteeing quality, following certain product standards or labor safety requirements.
	In practical terms, an official standardization organization could act as such a certifier.
	In our example, it verifies that logs coming from the forester meet certain quality guidelines. It doesn't own, create or handle any tokens itself.
  	It is also possible, that the sawmill creates its own certificate for equally usable goods.
	It can then utilize all products that have been certified to qualify to a certain level of similarity.

\end{enumerate}

\begin{table}[b]
\centering
\setlength{\tabcolsep}{4pt}
	\begin{tabular}{lcccc}
		\hline\noalign{\smallskip}
		\multirow{2}{*}{Roles}
		& \multicolumn{4}{c}{Actions} \\
		\noalign{\smallskip}
		\cline{2-5}
		\noalign{\smallskip}
		& Create & Consume & Split / Merge / Transfer & Certify \\
		\noalign{\smallskip}
		\hline
		\noalign{\smallskip}
		Resource Supplier  & \checkmark & & \checkmark & \\
		Producer &  \checkmark & \checkmark & \checkmark & \\
		Logistics / Retail & & & \checkmark & \\
		Consumer & & \checkmark & & \\
		Certifier & & & & \checkmark \\
		\hline\noalign{\smallskip}
	\end{tabular}
	\caption{Role definition and actions performed within the system}
	\label{table:roles}
\end{table}

\begin{figure*}[!ht]
	\centering
	\includegraphics[width=\textwidth, angle=0]{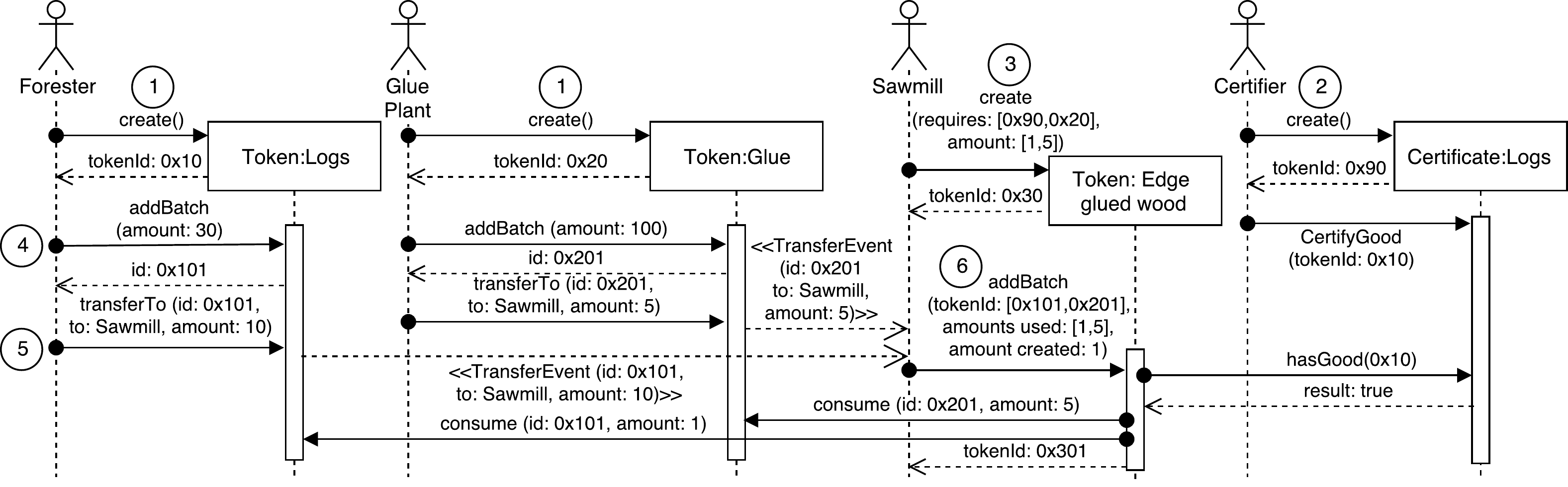}
	\caption{Example Use Case: After a Certifier approves a token that is used for defining required inputs in novel tokens}
	\label{figure:usecase}
\end{figure*}

\subsection{Traceability}
The physical production process is fully reflected on the blockchain. It works across multiple entities, as they cannot proceed without the correct inputs. Every step of the production process is accountable, and therefore traceability is not only provided for a single good, but also for its ingredients.
Depending on the interval in which new blocks are added to the blockchain, the block creation timestamp informs about the time and date a given product has been manufactured at.
As products are handled in batches, input information is available by batch. For example, a batch of edge glued wood could be traceable to a batch of 100 logs. In order to accomplish traceability for single goods, it should be declared as its own batch. However, fine-grained traceability comes with the cost of higher transaction counts.

\section{Implementation}
To achieve advanced traceability in supply chains, we propose a set of solidity contracts which are deployable on any EVM compatible blockchain. As ERC 20 tokens are not distinguishable and ERC 721 tokens do not permit defining minting conditions or measures of comparability between tokens, they do not satisfy our initial requirement of identifiability. Therefore, we suggest the representation of product batches as unique data structures which hold certain characteristics, such as the goods which have been consumed during the production process. Each batch of products corresponds to one token that holds unique features. Resource suppliers and manufacturers hold their own set of token contracts, one for each kind of product. To facilitate the deployment of novel token contracts, they can utilize a factory contract. While a token contract's ownership always remains with the initial deployer, the batches may be traded and change owners. Certifiers issue certificates to supply a measure of comparability between tokens.

To clarify our approach, we follow the example provided in our concept, but select a subset of processes for improved readability. The simplified example contains three parties who operate along a supply chain in six steps. 
The first entities involved are resource suppliers, namely a forester and glue plant who produce and sell wood and glue correspondingly, as presented in \mbox{Figure \ref{figure:usecase}}. The parties reflect production and trading processes along the supply chain following these subsequent steps:
\begin{enumerate}
	\item Token contracts are created that hold all produced batches of wood and glue that were produced by that specific forester and glue plant. As these goods do not involve any direct inputs, no additional products are required to add a batch of wood or glue in the contract.
	\item A certifier approves the forester's log tokens if the organization's measures apply. This step is essential for defining inputs for goods which need to comply with certain standards. For instance, when a  sawmill creates a board token contract, it may want to guarantee a high level of wood quality for all boards produced. Instead of specifying a particular wood contract, certificates can be used.
	\item The sawmill defines two required inputs, any logs that are certified by the certifier and the specific glue plant's glue. Through this mechanism, the sawmill is not bound to a single supplier but is enabled to either define a certificate to which consumed products need to comply or to directly specify supplier's good. 
	\item To add a batch of logs, the forester does not require any inputs, as s/he acts in the role of a resource producer. The created batch may be split thereafter, resulting in two or more new batches. Split operations are not visualized in Figure \ref{figure:usecase} to prevent introducing further complexity.
	\item The created batch of ten units is sent to the sawmill which is notified through an event triggered in the \textit{transfer} contract function. The glue plant performs corresponding steps.
	\item To add a batch of edge glued wood, the sawmill needs to define the wood and glue used during production. The \texttt{addbatch} function first checks if the defined wood is certified by querying the certifier's certificate contract. If the result is positive, wood is consumed, i.e. the wood token's \texttt{consume} function is called, reducing the overall amount of wood the forester's wood contract holds. The consumption of tokens is only possible for owned token batches which in our example is the case as the forester transfered the ownership in Step 5. As the glue token was defined directly without utilizing a certificate before, the token is reduced instantly.
\end{enumerate}	
As a result, the forester holds twenty units of wood, the glue plant owns fifty-nine units of glue while the sawmill holds nine units of wood, no glue and one unit of edge glued wood.
\subsection{Design Decisions}
Certifiers deploy their own contracts that hold certificates for multiple goods which possess similar characteristics. Alternatively, these organizations could issue signed certificates which are added to corresponding products' smart contracts. This approach enables observers to receive all certificates a good holds without querying a third contract. However, the certifier's signature has to be verified, adding overhead. When storing certified products in a dedicated certificate contract, the certificate has to be known before querying  for a specific product, while in the former approach this is not the case. Nevertheless, certificate contracts provide an important property for providing an ontology of products' similarity. Hereby, certificates can be declared as inputs for tokens rather than defining a specific good which would enforce a single supplier for sourcing. \note[m]{Revoke certificates - trust certifier}

\add[m]{For generating unique batch identifiers, resources, sender address and time are hashed using a SHA-3 hash function. While hashes bear the disadvantage of possible collisions, using a counter would require another storage operation every time a new batch is created. As the possibility of a collision within a single token contract with a byte length of 12 is in the measures of approximately  \(1 : 7.9*10^{18}\), we assume it to be small enough for relying on this mechanism to decrease operational gas costs.}

\subsection{Smart contract optimizations}
In the EVM, gas costs are assigned to each operation. Despite deploying new contracts, storing variables in the contract storage is the most expensive instruction~\cite{Wood2014}. To decrease operational costs, it is therefore desirable to minimize such costly procedures.
As a first measure, we use events rather than storage in case the data is not accessed within the contract (cf. Section \ref{subsec:events}).
Secondly, the word size in the EVM is 32 bytes long so that storing smaller types results in costs reflecting the full word size.
The solidity optimizer\footnote{http://solidity.readthedocs.io/en/develop/miscellaneous.html\#internals-the-optimizer} merges smaller variables to mitigate the involved expenses under certain conditions.
We exemplify this behavior in Figure \ref{figure:integer_comparison}a), using uint256 for storing input amounts results in a linear growth in gas costs with an increasing number of inputs. 
In contrast, when using uint32 we observe a lower slope but a step every eight inputs.
This is due to the fact that the optimizer concatenates up to eight variables in a single storage location before allocating a new one.
However, when storing only few variables, the overhead for merging variables in a single location exceeds its benefits.
This mechanism is only capable of concatenating types with the size of 8 bytes or multiple of it \cite{Garcia2017}.
For that reason, we conduct manual optimizations. 

\begin{figure}[h]
	\centering
	\includegraphics[width=0.49\textwidth, angle=0]{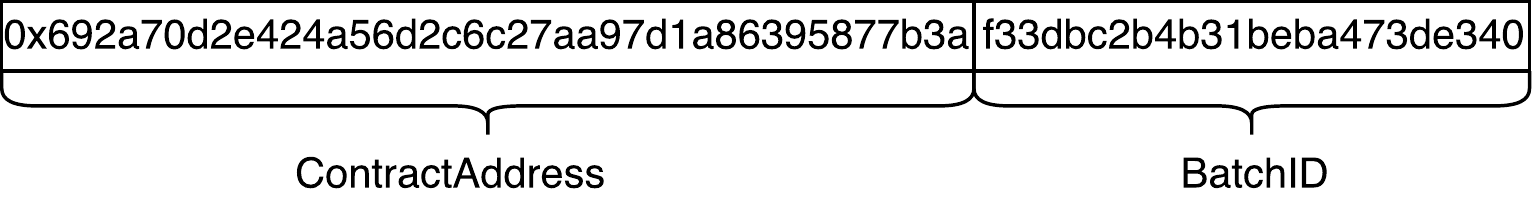}
	\caption{Contract addresses and batch identifiers are saved in a single 32 byte array to minimize storage operations}
	\label{figure:bytes32}
\end{figure}

When storing required inputs for creating tokens, we demand the inputs' contract address, batch identifiers and the amounts needed to create one product unit. Contract addresses in Ethereum are 40 hexadecimal digits long and therefore require 20 bytes for storage. While usually the remaining 12 bytes are padded with zeros, we use 12 byte long batch identifiers and concatenate them with the corresponding contract added, resulting in a single 32 byte write operation, as illustrated in Figure \ref{figure:bytes32}. The size of 24 hexadecimal digits provides sufficient space for storing unique identifiers and is derived from \mbox{SHA-3} hash in the contract. The concatenation and respective split operations are implemented using in-line assembly to provide a gas efficient calculation.

The use of smart contracts should be decoupled from user interaction. Adding a new product is therefore conducted through a factory contract which is called using Ethereum's JavaScript API called web3\footnote{http://web3js.readthedocs.io/en/1.0/index.html} and deploys a corresponding token contract on the blockchain. Hereby, the maintainability is increased, as front-end applications do not need to handle the contract code and following migrations are deployable through the factory contract.

\subsection{Events for analytics}
\label{subsec:events}
As the main target of our approach is to increase transparency in supply chains, we implemented a web-based tool to track provenance and manage tokens. Intuitively, batch inputs would be stored within their corresponding structs and pulled via getter methods. As provenance information is not required by contracts but only queried for analysis purposes, the token contract emits events rather than storing all information in order to decrease gas costs. Declaring topics when emitting events enables filtering and searching for attributes but are more expensive than using a raw data format. Even though events are not stored, they are reproducible, verifiable and easily observable due to the use of bloom filters for receiving relevant blocks in the blockchain \cite{Wood2014}.

We utilize the emitted events for providing traceability of products by recursively querying the inputs that have been declared for the token's creation. As the contract enforces the consumption of these inputs, the real world production process is reflected. Respectively, receiving all inputs in a recursive manner results in a tree of production inputs and unveils all resources used.

\section{Evaluation}
\label{sec:evaluation}
The prototypical implementation of a supply chain traceability system based on smart contracts raises various questions regarding deployment and maintenance costs as well as scalability. As such drawbacks are frequent issues when utilizing blockchain technologies~\cite{wuest2017}, we focus on these topics in the following sections for evaluating the proposed system's applicability.
\subsection{Gas costs}
\label{subsec:gascosts}
For analyzing the involved costs, we assume gas consumption as defined in Ethereum's yellow paper~\cite{Wood2014}. We refrain from converting measured gas costs to Ethereum's native currency or conventional currencies due to the lack of expressiveness. Such conversions would not only be subject to fluctuating exchange rates but also changing gas costs which depend on various factors such as current blocks' gas use and waiting time until a transaction is mined (cf. \ref{subsec:smartcontracts}).
In addition, the proposed approach is independent from the underlying blockchain and may be run for example in a permissioned manner. 

We differentiate between the costs for deploying contracts and execution costs. As participants create a contract for each type of product they offer, deployment costs are only relevant for novel products. The actual deployment costs depend on the amount of inputs declared, as they have to be stored in the contract. This information is required within the contract to ensure the right amount of goods has been consumed for creating a corresponding token, so that using events rather than storing the data is not applicable here. 

\begin{figure}
	\center
	\includegraphics[width=0.49\textwidth, angle=0]{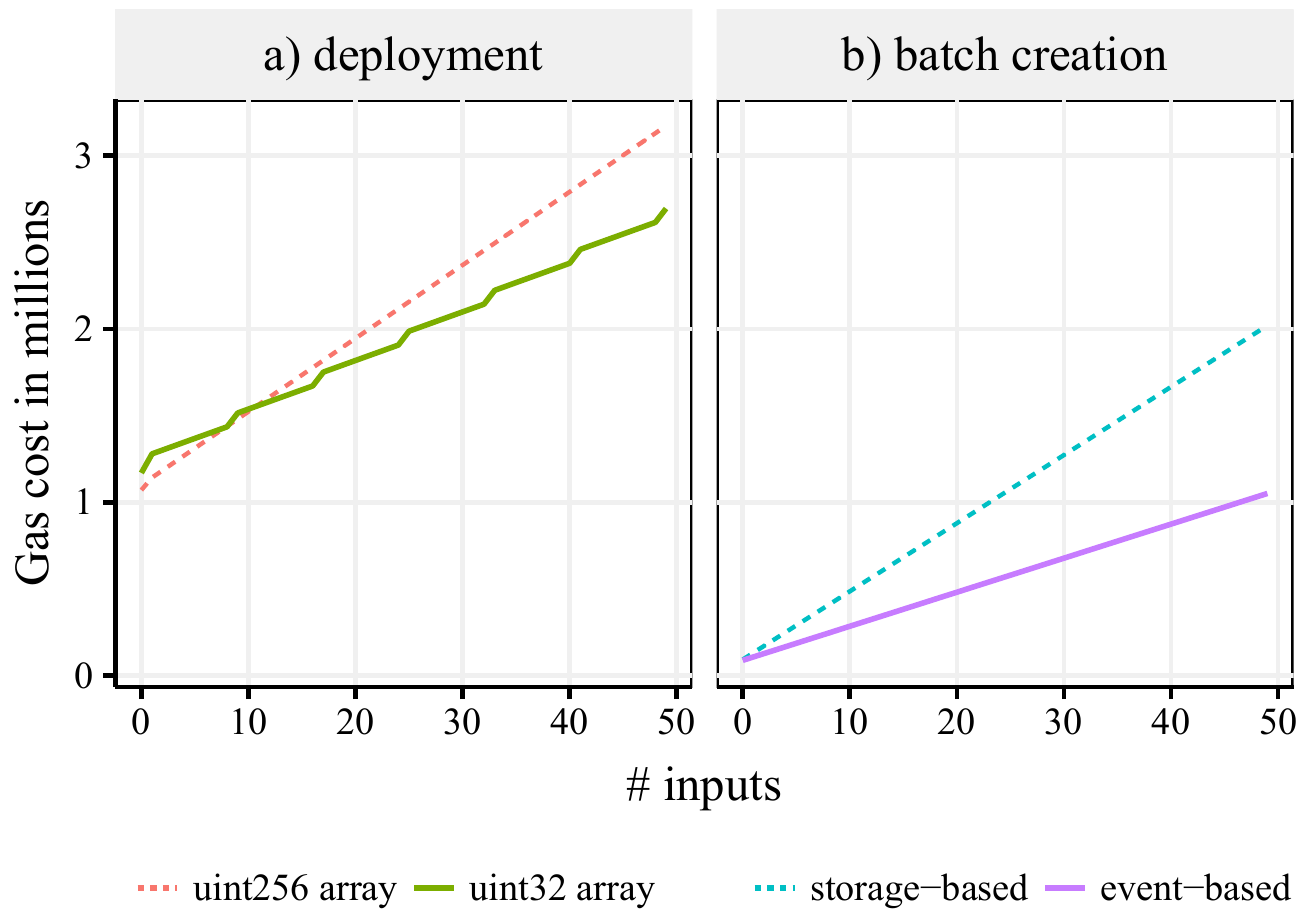}
	\caption{Gas costs for contract deployment and batch creation by implementation and number of inputs}
	\label{figure:integer_comparison}
\end{figure}

However, this is not the case for adding new batches. For obtaining comprehensive traceability information, emitted events are retrieved from a blockchain node using the web3 API. As this data is not required for smart contract operations, using events is sufficient so that transformation information is not stored within the contract, leading to decreased operational costs. Figure \ref{figure:integer_comparison}b) depicts a comparison of both approaches with respect to the amount of input goods. We observe a significantly steeper increase in gas costs when storing inputs in comparison to emitting events. 
While gas costs grow linearly with both approaches, we observe a factor of 39,340 gas for each added input in the classic storage approach but only 19,241 gas for using events, so just about half of the former.
The base costs for adding a batch with no input goods account for 92,756 gas.

To picture more complex scenarios, we create product tokens with two inputs on multiple tiers.
Consequently, the resulting relationships can be represented as a binary tree with its height relating to the level of tiers tracked in a supply chain.
We create batches with multiple tiers and compare the gas costs with a single batch that includes all inputs directly instead of using multiple tiers.
The resulting gas costs confirm our assumption that gas costs depend on the amount of vertices and egdes in the sourcing tree.
As a result, the overall gas costs do not depend on the amount of tiers considered in a supply chain but on the amount of sources.
This allows precise cost estimations for maintaining a traceability system.

\subsection{Scalability}
The system's scalability depends on a multitude of variables. Most importantly, the decision of utilizing a public or permissioned ledger depends on the desired scope, accessibility and privacy considerations. In case a public ledger is preferred, the scalability highly depends on the involved gas costs (cf. Section~\ref{subsec:gascosts}).
The scalability of a permissioned ledger varies with its implementation of block size limitations, consensus algorithm and mining time, determining attributes such as throughput and ledger size. As our implementation is agnostic to the underlaying blockchain, no final assessment is feasible here. Tests using the Ethereum client geth\footnote{https://github.com/ethereum/go-ethereum} unveiled a plethora of aspects that influence ledger growth. For instance, the file system's block size affects the ledger size as files created by the client may fill only fractions of it, leading to increased disk space consumption.

Varying batch sizes influence the systems scalability in terms of throughput, ledger size and potential gas costs. Consequently, small batch sizes, as required for tracing single goods, negatively affect the system's performance.


\section{Discussion and Outlook}

\subsection{Incentives}
Employing supply chain traceability on a blockchain implies transparency to an extent that raises questions regarding participants' willingness to disclose otherwise confidential information. The degree to which information is published depends on the underlying blockchain. Using a public ledger entails unveiling information to a large audience, while utilizing a permissioned ledger restricts the number of participants but may still involve revealing information to competitors. Therefore, the proposed solution is only applicable for use cases in which the desired transparency requirements exceed the threat of losing competitive advantage by disclosing internal information. This may be the case if customers have a high interest in resource provenance or an increased quality level in an environment with limited trust.

The main incentive for participating in a traceability system is to increase customer's trust in the delivered product.
For example, customers buying food are willing to buy larger quantities and pay higher price if they are enabled to reconstruct a product's provenance~\cite{Choe2009}.
With increasingly complex supply chains, guaranteeing high quality has become a difficult endeavor, generating the need for high transparency.
\newpage
\subsection{Outlook}
The proposed solution should be understood as a foundation for blockchain-based supply chain management systems that can be built upon. In its current stage it does not yet cover functionalities that may be required in a real-world setting. We anticipate the following features to be of interest:

\begin{itemize}
	\item \textit{Payment for goods}: In our solution, payments are not considered. It is questionable whether it would be desirable for a business to open its books on how much has been paid for the acquisition of inputs for a manufacturing process. If purchases were public however, inefficiencies of the market would be transparent and could be exploited.

	\item \textit{Shrinkage}: If a good is damaged or lost, the system currently has no suitable mechanism to capture such an event, other than letting the affected business simply consume the respective tokens in order to remove them from the supply chain.

	\item \textit{Ownership vs. possession}: It may be desirable to let a logistics firm possess goods that are owned by another party. This raises additional questions if such a logistics firm suffers from shrinkage.

	\item \textit{Packaging}: There are scenarios in which goods may need to be packaged together and extracted at a later stage. With the current solution, a packaging process can be implemented similarly to a manufacturing process, but the extraction of the original goods is not possible, inhibiting traceability.
\end{itemize}

\section{Conclusion}
We have proposed a blockchain-based  supply chain management system that enables  tracing and tracking goods including their transformation in the production process using smart contracts. The system provides comprehensive provenance information by projecting product compositions onto the blockchain in the form of tokens. Hereby, shortcomings of current traceability systems regarding isolated data storage and lacking transformation information are tackled. Defining compositions for creating products and enforcing them using smart contract enables the documentation of consumed resources in the production process. Through this mechanism, products are not only traceable from production to retail but starting from resource exploitation. As a result, transparency is generated along the supply chain, providing comprehensible production information.

\bibliographystyle{IEEEtran}

\begin{thebibliography}{10}
\providecommand{\url}[1]{#1}
\csname url@samestyle\endcsname
\providecommand{\newblock}{\relax}
\providecommand{\bibinfo}[2]{#2}
\providecommand{\BIBentrySTDinterwordspacing}{\spaceskip=0pt\relax}
\providecommand{\BIBentryALTinterwordstretchfactor}{4}
\providecommand{\BIBentryALTinterwordspacing}{\spaceskip=\fontdimen2\font plus
\BIBentryALTinterwordstretchfactor\fontdimen3\font minus
  \fontdimen4\font\relax}
\providecommand{\BIBforeignlanguage}[2]{{%
\expandafter\ifx\csname l@#1\endcsname\relax
\typeout{** WARNING: IEEEtran.bst: No hyphenation pattern has been}%
\typeout{** loaded for the language `#1'. Using the pattern for}%
\typeout{** the default language instead.}%
\else
\language=\csname l@#1\endcsname
\fi
#2}}
\providecommand{\BIBdecl}{\relax}
\BIBdecl

\bibitem{Dabbene2014}
F.~Dabbene, P.~Gay, and C.~Tortia, ``Traceability issues in food supply chain
  management: A review,'' \emph{Biosystems Engineering}, vol. 120, pp. 65 --
  80, 2014.

\bibitem{Gualandris2015}
J.~Gualandris, R.~D. Klassen, S.~Vachon, and M.~Kalchschmidt, ``Sustainable
  evaluation and verification in supply chains: Aligning and leveraging
  accountability to stakeholders,'' \emph{Journal of Operations Management},
  vol.~38, pp. 1 -- 13, 2015.

\bibitem{eu178}
\BIBentryALTinterwordspacing
{European Parliament }, ``Regulation (ec) no 178/2002 of the european
  parliament and of the council of 28 january 2002 laying down the general
  principles and requirements of food law, establishing the european food
  safety authority and laying down procedures in matters of food safety,''
  2002. [Online]. Available:
  \url{http://eur-lex.europa.eu/legal-content/EN/TXT/PDF/?uri=CELEX:32002R0178\&from=EN}
\BIBentrySTDinterwordspacing

\bibitem{Appelhanz2016}
S.~Appelhanz, V.-S. Osburg, W.~Toporowski, and M.~Schumann, ``Traceability
  system for capturing, processing and providing consumer-relevant information
  about wood products: system solution and its economic feasibility,''
  \emph{Journal of Cleaner Production}, vol. 110, pp. 132 -- 148, 2016, special
  Volume: Improved resource efficiency and cascading utilisation of renewable
  materials.

\bibitem{Abeyratne2016}
S.~A. Abeyratne and R.~P. Monfared, ``{Blockchain ready manufacturing supply
  chain using distributed ledger},'' \emph{International Journal of Research in
  Engineering and Technology}, vol.~05, September 2016.

\bibitem{korpela2017}
K.~Korpela, J.~Hallikas, and T.~Dahlberg, ``{Digital Supply Chain
  Transformation toward Blockchain Integration},'' in \emph{Proceedings of the
  50th Hawaii International Conference on System Sciences}, 2017.

\bibitem{glaser2017}
F.~Glaser, ``{Pervasive Decentralisation of Digital Infrastructures: A
  Framework for Blockchain enabled System and Use Case Analysis},'' in
  \emph{Proceedings of the 50th Hawaii International Conference on System
  Sciences}, 2017.

\bibitem{wuest2017}
K.~W\"ust and A.~Gervais, ``{Do you need a Blockchain?}'' Cryptology ePrint
  Archive, Report 2017/375, 2017.

\bibitem{Kim2016}
H.~M. Kim and M.~Laskowski, ``{Towards an Ontology-Driven Blockchain Design for
  Supply Chain Provenance},'' \emph{SSRN Electronic Journal}, August 2016.

\bibitem{Toyoda2017}
K.~Toyoda, P.~T. Mathiopoulos, I.~Sasase, and T.~Ohtsuki, ``{A Novel
  Blockchain-Based Product Ownership Management System (POMS) for
  Anti-Counterfeits in The Post Supply Chain},'' \emph{IEEE Access}, vol.~PP,
  no.~99, 2017.

\bibitem{Hackius2017}
N.~Hackius and M.~Petersen, ``\BIBforeignlanguage{en_US}{{Blockchain in
  Logistics and Supply Chain: Trick or Treat?}}'' in
  \emph{\BIBforeignlanguage{en_US}{Proceedings of the Hamburg International
  Conference of Logistics (HICL)}}, October 2017, pp. 3--18.

\bibitem{BECHINI2008342}
A.~Bechini, M.~G. Cimino, F.~Marcelloni, and A.~Tomasi, ``Patterns and
  technologies for enabling supply chain traceability through collaborative
  e-business,'' \emph{Information and Software Technology}, vol.~50, no.~4, pp.
  342 -- 359, 2008.

\bibitem{Gunasekaran2004}
A.~Gunasekaran and E.~Ngai, ``Information systems in supply chain integration
  and management,'' \emph{European Journal of Operational Research}, vol. 159,
  no.~2, pp. 269 -- 295, 2004.

\bibitem{Helo2005}
P.~Helo and B.~Szekely, ``Logistics information systems: An analysis of
  software solutions for supply chain co‐ordination,'' \emph{Industrial
  Management \& Data Systems}, vol. 105, no.~1, pp. 5--18, 2005.

\bibitem{Tse2011}
Y.~K. Tse and K.~H. Tan, ``Managing product quality risk in a multi-tier global
  supply chain,'' \emph{International Journal of Production Research}, vol.~49,
  no.~1, pp. 139--158, 2011.

\bibitem{Costa2013}
C.~Costa, F.~Antonucci, F.~Pallottino, J.~Aguzzi, D.~Sarri{\'a}, and
  P.~Menesatti, ``{A Review on Agri-food Supply Chain Traceability by Means of
  RFID Technology},'' \emph{Food and Bioprocess Technology}, vol.~6, no.~2, pp.
  353--366, February 2013.

\bibitem{Yuan2016}
Y.~Yuan and F.~Y. Wang, ``{Towards Blockchain-based Intelligent Transportation
  Systems},'' in \emph{2016 IEEE 19th International Conference on Intelligent
  Transportation Systems (ITSC)}, November 2016, pp. 2663--2668.

\bibitem{szabo1994smart}
N.~Szabo, ``Smart contracts,'' 1994.

\bibitem{Buterin2014}
V.~Buterin, ``{A next-generation smart contract and decentralized application
  platform},'' \emph{Ethereum Project White Paper}, 2014.

\bibitem{Wood2014}
G.~Wood, ``{Ethereum: a secure decentralised generalised transaction ledger},''
  \emph{Ethereum Project Yellow Paper}, 2014.

\bibitem{vanDorp2003}
C.~Van~Dorp, ``Tracking and tracing business cases: Incidents, accidents and
  opportunities,'' \emph{Proceedings of EFITA Conference}, pp. 601--606, 2003.

\bibitem{Feng2017}
F.~Tian, ``{A Supply Chain Traceability System for Food Safety Based on HACCP,
  Blockchain \& Internet of Things},'' in \emph{2017 International Conference
  on Service Systems and Service Management}, 2017.

\bibitem{feng2016}
{F. Tian}, ``{An Agri-food Supply Chain Traceability System for China Based on
  RFID \& Blockchain Technology},'' in \emph{2016 13th International Conference
  on Service Systems and Service Management (ICSSSM)}, June 2016.

\bibitem{Garcia2017}
L.~Garc{\'i}a-Ba{\~{n}}uelos, A.~Ponomarev, M.~Dumas, and I.~Weber,
  ``{Optimized Execution of Business Processes on Blockchain},'' in
  \emph{Business Process Management}, J.~Carmona, G.~Engels, and A.~Kumar,
  Eds.\hskip 1em plus 0.5em minus 0.4em\relax Springer International
  Publishing, 2017, pp. 130--146.

\bibitem{Choe2009}
Y.~C. Choe, J.~Park, M.~Chung, and J.~Moon, ``Effect of the food traceability
  system for building trust: Price premium and buying behavior,''
  \emph{Information Systems Frontiers}, vol.~11, no.~2, pp. 167--179, April
  2009.

\end{thebibliography}

\end{document}